\documentclass{article}
\usepackage{graphicx} 
\usepackage{graphics}
\usepackage{amsthm}
\usepackage{amsmath}
\usepackage{mathrsfs}
\usepackage{graphicx}
\usepackage{floatrow}
\usepackage{caption}
\usepackage{subcaption}
\usepackage{soul}
\usepackage{mathtools}
\usepackage[margin=2.5cm]{geometry}
\DeclarePairedDelimiter\bra{\langle}{\rvert}
\DeclarePairedDelimiter\ket{\lvert}{\rangle}
\DeclarePairedDelimiterX\braket[2]{\langle}{\rangle}{#1\,\delimsize\vert\,\mathopen{}#2}
\usepackage{mathrsfs}
\usepackage{authblk}
\title{Impact of Polarization Distinguishability on Coherence in QIUP}

\author[1]{Gaytri Arya}
\author[1]{Paolo Bianchini}
\author[1,2]{Alberto Diaspro}

\affil[1]{Nanoscopy, Istituto Italiano di Tecnologia, Genoa, Italy}
\affil[2]{DIFILAB Department of Physics, University of Genoa, Genoa, Italy}

\begin{document}
\maketitle

\begin{abstract}
Quantum imaging with undetected photons (QIUP) is a promising technique for non-invasive imaging, particularly at wavelengths where detectors are inefficient or unavailable. Recent studies have investigated the role of polarization in QIUP schemes. In this work, we examine whether post-generation optical elements can recover interference when the signal photons are distinguishable in polarization at the source. Specifically, we employ a Mach-Zehnder interferometer after the photon pairs are generated from two nonlinear crystals. Our results indicate that polarization distinguishability at the time of generation leads to a loss of interference that cannot be restored by post-generation optical arrangements. This highlights the fundamental importance of indistinguishability at the source for maintaining coherence in QIUP systems.  
\end{abstract}

\section{Introduction}
Quantum imaging techniques offer several advantages over classical methods, including improved resolution, reduced noise, and minimal photon interaction with the sample. Among these, quantum imaging with undetected photons (QIUP) has emerged as a particularly promising approach. \cite{gilaberte2019perspectives} Based on the principle of induced coherence without induced emission, QIUP leverages biphotons produced via spontaneous parametric down-conversion (SPDC), where the idler photons interact with the sample while the corresponding signal photons, undetected by the sample, carry the relevant information \cite{zou1991induced,lemos2014quantum}. This imaging modality has been extensively studied both theoretically and experimentally to characterize key performance parameters such as resolution and field of view, which are critical for optimizing its imaging capabilities \cite{pearce2023practical,buzas2020biological,hochrainer2022quantum}. More recently, QIUP has been extended to probe polarization-dependent properties, enabling the investigation of polarization-sensitive or anisotropic samples. \\
A central feature of quantum imaging with undetected photons (QIUP) lies in the indistinguishability of the idler photons generated by the two sources (typically nonlinear crystals). Beyond precise optical alignment, factors such as transmission losses and polarization mismatches critically influence the degree of indistinguishability between the idler paths \cite{lahiri2017partial}. In polarization-based QIUP schemes, it is essential to ensure that the signal-idler pairs generated in the second source remain indistinguishable even when they are orthogonally polarized at the point of generation \cite{fuenzalida2024quantum,lahiri2021characterizing}. This typically necessitates the use of two separate nonlinear crystals to generate both horizontally and vertically polarized photon pairs. \\
In this work, we propose an alternative configuration that replaces one of the crystals with a modified Mach-Zehnder interferometer (MZI). The MZI includes a half-wave plate (HWP) in one arm, designed to produce orthogonally polarized photons while erasing which-path information. However, our theoretical analysis reveals that polarization-induced distinguishability in the idler photons—originating from their generation in separate optical paths—cannot be fully mitigated by post-generation optical adjustments. This residual distinguishability limits the effectiveness of the proposed substitution in replicating the behavior of conventional dual-crystal setups.   

 
\section{Mathematical Framework} 
\begin{figure}[h]
    \centering
    \includegraphics[width=1\linewidth]{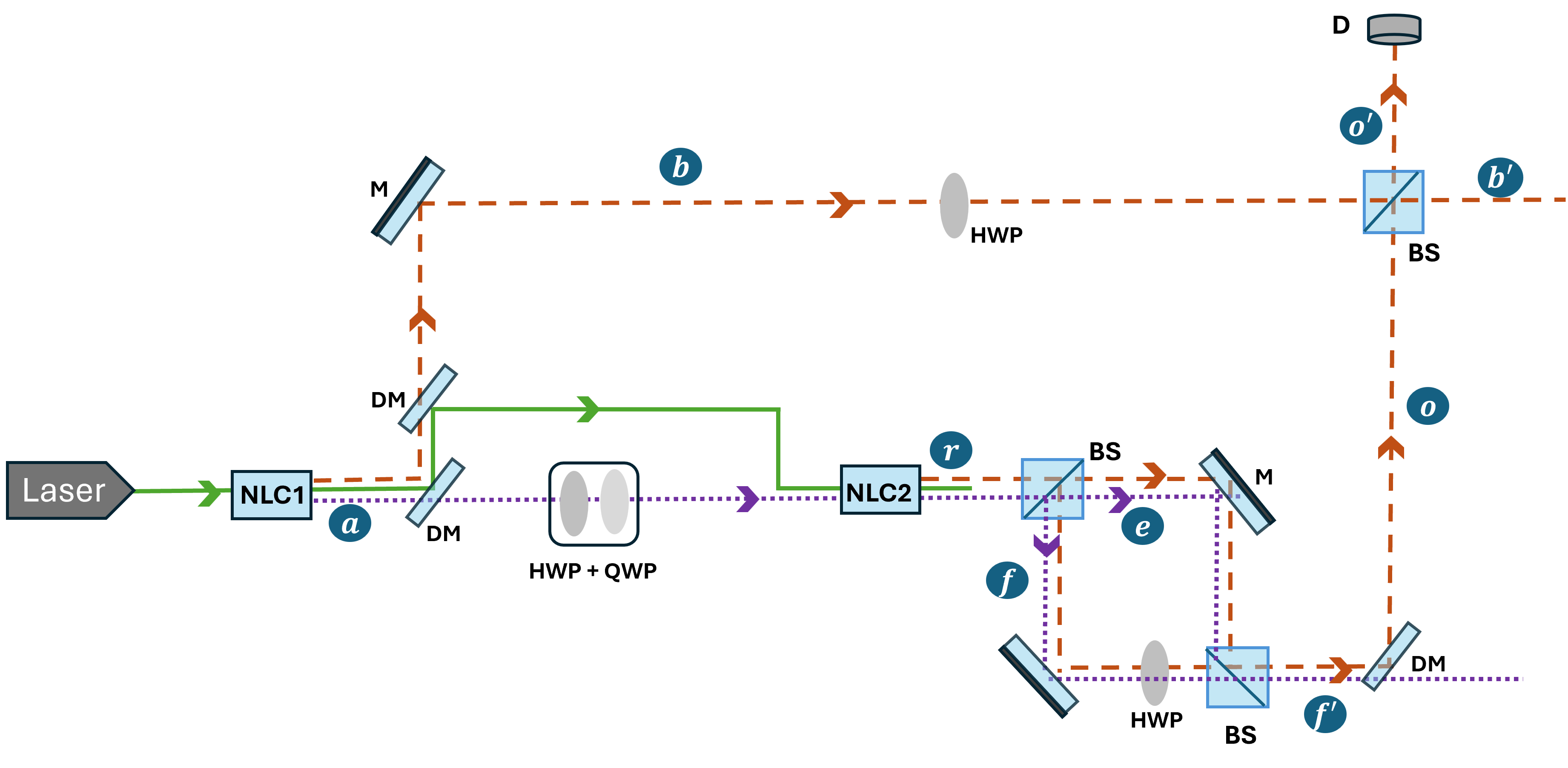}
    \caption{Scheme to explore polarization in QIUP using a MZI after the second crystal. Notations: NLC Non-linear crystal, DM Dichroic mirror, HWP Half-wave plate, QWP Quarter-wave plate, BS Beam splitter, M Mirror, D Detector. Green solid line represents pump laser, rust dashed line indicates signal photons and the violet dotted line indicates idler photons.}
    \label{fig:enter-label}
\end{figure}
In this section, we establish a mathematical proof of the inability to compensate for the distinguishability of different polarizations of idler photons in the QIUP scheme. This scheme is composed of two crystals as our sources (NLC1 and NLC2), which are exactly similar to each other and generate vertically polarized signal-idler photons. We are assuming that the frequencies of signal and idler photons are quite apart from each other. In the beginning, when two crystals produce vertical photons in their respective spatial modes, initial wavefunction is described as
\begin{equation}
    \ket{\psi_{initial}}=  \ket{V_{I1}}_a \ket{V_{S1}}_a +  \ket{V_{I2}}_r \ket{V_{S2}}_r
\end{equation}  
The dichroic mirror separates the signal photon in path \textit{b}, whereas the combination of a half-wave plate and a quarter-wave plate modifies the idler photon state. 
The following matrices depict the operation of HWP and QWP with $\textit{h}$ and $\textit{q}$ being the angle of the fast axis with x-axis of HWP and QWP, respectively \cite{james2001measurement},
\begin{align}
    \Hat{U}_{QWP} = &	\begin{pmatrix}
\iota-\cos{[2q]} & \sin{[2q]}\\
\sin{[2q]} & \iota+\cos{[2q]} \end{pmatrix}, \\
    \Hat{U}_{HWP} = &	\begin{pmatrix}
\cos{[2h]} & -\sin{[2h]}\\
-\sin{[2h]} & -\cos{[2h]} \end{pmatrix}.
\end{align}
With a combination of HWP and QWP in idler path, and an HWP in signal path \textit{b}, using the standard wavefunction notation, the state of the  resulting system before the MZI is given as 
\begin{equation}
    \ket{\psi}= (\alpha_1 \ket{H_{I1}}_r +  \beta_1 \exp{(\iota \gamma)} \ket{V_{I1}}_r) (\alpha_2 \ket{H_{S1}}_b +  \beta_2 \ket{V_{S1}}_b) + \exp{(\iota \phi)} \ket{V_{I2}}_r \ket{V_{S2}}_r
\end{equation}
Here, $|\alpha_1|^2$ and $|\beta_1|^2$ are horizontal and vertical probabilities, whereas $\gamma$ is the phase difference between the horizontal and vertical components of the idler photons coming from NLC1. On the other hand,  $|\alpha_2|^2$ and $|\beta_2|^2$ are the probabilities of finding the signal photons in horizontal and vertical polarizations. The phase $\phi$ refers to the interferometric phase introduced between two sources. Since the two crystals are exactly identical and for the case of perfect alignment, the idler photons of NLC2 are indistinguishable from those of NLC1. But our sources are generating only vertical photons. Hence, this condition of indistinguishability applies only for vertical photons in path \textit{r}.
\begin{equation}
    \begin{split}
    &   \ket{V_{I1}}_r \rightarrow \ket{V_{I}}_r \\
  &   \ket{V_{I2}}_r \rightarrow \ket{V_{I}}_r \\
  &  \ket{V_{S1}}_{b} \rightarrow \ket{V_{S}}_{b}  \\
  &   \ket{V_{S2}}_{r} \rightarrow \ket{V_{S}}_{r}  \\
    \end{split}
\end{equation}
The state takes the form
\begin{equation}
    \ket{\psi}= (\alpha_1 \ket{H_{I1}}_r +  \beta_1 \exp{(\iota \gamma)} \ket{V_{I}}_r) (\ket{H_{S1}}_b +  \beta_2 \ket{V_{S}}_b) + \exp{(\iota \phi)} \ket{V_{I}}_r \ket{V_{S}}_r
\end{equation}
The BS in path $r$ splits the stream of photons into the paths \textit{e} and \textit{f},
\begin{equation}
\label{BS}
    \ket{Y}_r \rightarrow \frac{\ket{Y}_e + \iota \ket{Y}_f}{\sqrt{2}}: Y=H,V.
\end{equation}
The state of the system becomes
\begin{equation}
\begin{split}
    \ket{\psi} = &   (\alpha_2 \ket{H_{S1}}_b + \beta_2 \ket{V_S}_b) \Big(\alpha_1 \frac{\ket{H_{I1}}_e + \iota \ket{H_{I1}}_f}{\sqrt{2}} + \beta_1 \exp{(\iota \gamma) \frac{\ket{V_I}_e + \iota \ket{V_I}_f}{\sqrt{2}}}\Big) \\
    & + \exp{(\iota \phi)} \Big(\frac{\ket{V_S}_e + \iota \ket{V_S}_f}{\sqrt{2}}\Big) \Big(\frac{\ket{V_I}_e + \iota \ket{V_I}_f}{\sqrt{2}}\Big)
    \end{split}
\end{equation}
The HWP in path $f$ modifies the idler and signal photons as follows
\begin{equation}
    \begin{split}
        & \ket{H_{I1}}_f \rightarrow \cos{[2 \theta]} \ket{H_{I1}}_f - \sin{[2 \theta]} \ket{V_{I1}}_f \\
      & \ket{V_I}_f \rightarrow - \sin{[2 \theta]} \ket{H_I}_f - \cos{[2 \theta]} \ket{V_I}_f \\
          & \ket{V_S}_f \rightarrow - \sin{[2 \theta]} \ket{H_S}_f - \cos{[2 \theta]} \ket{V_S}_f \\
    \end{split}
\end{equation}

\begin{equation}
    \begin{split}
        \ket{\psi} = & \frac{1}{\sqrt{2}}  (\alpha_2 \ket{H_{S1}}_b + \beta_2 \ket{V_S}_b) \big( \alpha_1 \ket{H_{I1}}_e + \beta_1 \exp{(\iota \gamma)} \ket{V_I}_e \big) \\
        & + \frac{ \beta_1 \exp{(\iota \gamma)}}{\sqrt{2}}  (\alpha_2 \ket{H_{S1}}_b + \beta_2 \ket{V_S}_b) \Big(- \iota \sin{[2 \theta] \ket{H_I}_f - \iota \cos{[2 \theta]} \ket{V_I}_f}\Big) \\
        & + \frac{ \alpha_1}{\sqrt{2}}  (\alpha_2 \ket{H_{S1}}_b + \beta_2 \ket{V_S}_b) \Big( \iota \cos{[2 \theta ]} \ket{H_{I1}}_f - \iota \sin{[2 \theta]} \ket{V_{I1}}_f \Big) \\
       & + \frac{\exp{(\iota \phi)}}{2} \Big(\ket{V_S}_e - \iota \sin{[2 \theta]} \ket{H_S}_f - \iota \cos{[2 \theta]} \ket{V_S}_f \Big)\Big(\ket{V_I}_e - \iota \sin{[2 \theta]} \ket{H_I}_f - \iota \cos{[2 \theta]} \ket{V_I}_f \Big)
    \end{split}
\end{equation}
Applying a similar action as given in Eq.\eqref{BS}  of the second BS at the exit of the MZI (splitting path \textit{e} into $\textit{e}^{'}$ and  $f^{'}$, and similar for \textit{f}). Afterwards, with the use of DM, we separated out the signals from path $f^{'}$ into the path \textit{o}, i.e.
\begin{equation}
   \begin{split}
      &  \ket{H_S}_{f^{'}} \rightarrow \ket{H_S}_o \\
      &  \ket{V_S}_{f^{'}} \rightarrow \ket{V_S}_o 
   \end{split}
\end{equation}
And lastly, BS facilitates the interference of the signal photons from paths $o$ and $b$, and we can measure in either path $o^{'}$ or $b^{'}$. 
We fix the detection in path $o^{'}$; therefore, the wavefunction is conditioned on the signal photons propagating along path $o^{'}$ is given as
\begin{equation}
    \ket{\psi}_{final} = \ket{\psi_1} + \ket{\psi_2} 
\end{equation}
\begin{equation}
   \begin{split}
        \ket{\psi_1} =   \ket{H_S}_{o^{'}}  & \Bigg[ \ket{H_I}_{f^{'}} \Big( \frac{-  \exp{(\iota \phi)} \sin^2[2 \theta] }{4\sqrt{2}} + \frac{ \alpha_2 \beta_1 \exp{(\iota \gamma)} \sin{[2 \theta]}}{2\sqrt{2}}  \Big) + \ket{H_{I1}}_{f^{'}} \frac{ \alpha_1 \alpha_2 (\cos{[2 \theta]}-1)}{ 2 \sqrt{2}} \\
        & + \ket{V_I}_{f^{'}} \Big(\frac{ \exp{(\iota \phi)} \sin{[ 2 \theta]} (1 - \cos{[2 \theta]})}{4 \sqrt{2}} + \frac{ \alpha_2 \beta_1 (\exp{(\iota \gamma)} \cos{[2 \theta]}-1)}{2 \sqrt{2}} \Big)  + \ket{V_{I1}}_{f^{'}} \frac{ \alpha_1 \alpha_2 \sin{[2 \theta] }}{ 2\sqrt{2}} \\
        & + \ket{H_I}_{e^{'}} \Big( \frac{-\iota \exp{(\iota \phi)} \sin^2{[2 \theta]}}{4 \sqrt{2}} - \frac{ \alpha_2 \beta_1 \exp{(\iota \phi)}}{2 \sqrt{2}} \Big) + \ket{H_{I1}}_{e^{'}} \frac{\iota  \alpha_1 \alpha_2 (1+ \cos{[2 \theta]})}{2 \sqrt{2}}  \\
        & +\ket{V_I}_{e^{'}} \Big(\frac{-\iota 
        \exp{(\iota \phi)} \sin[2\theta] (1+\cos{[2 \theta]})}{4\sqrt{2}} + \frac{\iota  \alpha_2 \beta_1 (\exp{(\iota \gamma)} \cos{[2 \theta]}+1)}{2 \sqrt{2}} \Big) + \ket{V_{I1}}_{e^{'}} \frac{\iota \alpha_1 \alpha_2 \sin{[2 \theta]}}{2 \sqrt{2}} \Bigg] \\
   \end{split}
\end{equation}

\begin{equation}
    \begin{split}
        \ket{\psi_2} = \ket{V_S}_{o^{'}}  & \Bigg[  \ket{H_I}_{f^{'}} \Big( \frac{ \exp{(\iota \phi )}\sin{[2 \theta]} (1-\cos[2 \theta])}{4 \sqrt{2}} + \frac{ \beta_1 \beta_2 \exp{(\iota \gamma)} \sin{[2 \theta]}}{2 \sqrt{2}} \Big) + \ket{H_{I1}}_{f^{'}} \frac{ \alpha_1 \beta_2 (\cos{[2 \theta]}-1)}{2 \sqrt{2}}  \\
       &  + \ket{V_I}_{f^{'}} \Big(\frac{-\exp{(\iota \phi)} (1-\cos{[2 \theta]})^2}{4 \sqrt{2}} + \frac{ \beta_1 \beta_2 (\exp{(\iota \gamma)} \cos{[2 \theta]}-1)} {2 \sqrt{2}} \Big) + \ket{V_{I1}}_{f^{'}} \frac{ \alpha_1 \beta_2 \sin{[2 \theta]}}{2 \sqrt{2}}  \\
       & + \ket{H_I}_{e^{'}} \Big( \frac{\iota \exp{(\iota \phi)} \sin{[2 \theta]} (1-\cos{[2 \theta]})}{4 \sqrt{2}} + \frac{\iota \beta_1 \beta_2 \exp{(\iota \gamma)} \sin[2 \theta] }{ 2\sqrt{2}} \Big) + \ket{H_{I1}}_{e^{'}} \frac{ \alpha_1 \beta_2 (\iota \cos{[2 \theta]}-1)}{2 \sqrt{2}} \\
     & + \ket{V_I}_{e^{'}} \Big(\frac{\iota \exp{(\iota \phi)} (1-\cos^2{[2 \theta]})}{4 \sqrt{2}} + \frac{\iota \beta_1 \beta_2 (\exp{(\iota \gamma)} \cos{[2 \theta]}+1)}{2 \sqrt{2}}  \Big)+ \ket{V_{I1}}_{e^{'}} \frac{\iota \alpha_1 \beta_2 \sin{[2 \theta]} }{2 \sqrt{2}} \Bigg]\\
    \end{split}
\end{equation}

The horizontal and vertical counts in path $o^{'}$ can be evaluated as 
\begin{equation}
    \begin{split}
        & \langle N_H \rangle = \bra{\psi}\hat{a}_{H_{o^{'}}}^{\dagger} \hat{a}_{H_{o^{'}}}\ket{\psi} \\
        &\langle N_V \rangle = \bra{\psi}\hat{a}_{V_{o^{'}}}^{\dagger} \hat{a}_{V_{o^{'}}}\ket{\psi} \\
    \end{split}
\end{equation}
Here, the $\hat{a}_{H_{o^{'}}}^{\dagger}$ and $\hat{a}_{H_{o^{'}}}$ are the creation and annhilation operator on horizontal photons in path $o^{'}$. When the HWP in the MZI is at $45^o$, it transforms the incoming vertical photons into horizontal, and vice-a-versa. Therefore, this orientation would result in both horizontal and vertical idler photons, and classically we expect that those idler photons are not differentiable from each other. But this is not the case, as the horizontal photons which were distinguishable at one point (after the crystal), would remain distinguishable in spite of the post arrangements of the optical components. For $\beta_2 =1$, the horizontal and vertical counts are obtained as
\begin{equation}
\label{Nh}
  \langle N_H \rangle= \frac{1}{16}\bigg(8 - 3 \beta_1^2 + \beta_1 (\sin{[\gamma-\phi]}-\cos{[\gamma-\phi]}) -2  \beta_1 \cos{[\phi]} \bigg)    
\end{equation}
\begin{equation}
\label{Nv}
 \langle N_V \rangle = \frac{1}{16}\bigg(5 + 2  \beta_1 \big(\cos{[\gamma-\phi]} + \cos{[\phi]} \big) \bigg) 
\end{equation}
If we want to determine the density matrix elements of the idler beam ($\alpha_1$, $\beta_1$ and $\gamma$), clearly we cannot directly evaluate $\alpha_1$ as both the horizontal and vertical counts contain $\beta_1$ information. This means we can obtain information about the vertical component of the idler photons, as the indistinguishability condition was fulfilled only for the vertical photons.
Even for a precise estimation of $\beta_1$, we need to calibrate the system such that we first set $\gamma$ to zero and then vary the phase. With this strategy, the visibility for the vertical counts comes out 
\begin{equation}
   \nu_V = \frac{4 \beta_1}{5},
\end{equation}
which we determined from the visibility formula 
\begin{equation}
    \nu_V = \frac{ \langle N_V \rangle_{max}- \langle N_V \rangle_{min}}{ \langle N_V \rangle_{max} +  \langle N_V \rangle_{min}}.
\end{equation}
For the final step of calibration, we note the angle $\phi$ corresponding to the maximum vertical counts ($\langle N_V \rangle_{\text{max}}$). For the unknown system, $\beta_1$ and $\gamma$ can then be determined by fitting the measured horizontal and vertical counts using Eqs.~\eqref{Nh} and \eqref{Nv}. This fitting enables the precise determination of both the vertical probability amplitude and the phase difference. Although the horizontal component of the idler photons ($\alpha_1$) cannot be directly accessed, in a non-lossy system, $\alpha_1$ can be inferred from the normalization condition $\alpha_1^2 + \beta_1^2 = 1$. The accuracy of this method is contingent on the assumptions of negligible transmission losses and low experimental noise. 

\section{Conclusion}
The QIUP (Quantum Imaging with Undetected Photons) scheme has attracted significant interest due to its potential advantages across various fields. A central requirement of the scheme is the indistinguishability of the idler photons generated by two SPDC sources. In this study, we investigated the role of polarization in influencing this indistinguishability. Specifically, we examined a setup involving two vertically polarized SPDC sources and explored whether a Mach-Zehnder interferometer (MZI) placed after the sources could erase the distinguishability of the horizontally polarized photons. Our analysis, based on the observed horizontal and vertical polarization counts, indicated that only information related to the vertical polarization component was accessible. This suggests that once polarization-based distinguishability is introduced at the generation stage, it cannot be effectively mitigated through post-generation optical arrangements alone. While these results point toward a limitation of post-processing in restoring indistinguishability, further studies may help clarify the boundaries and potential exceptions of this behaviour.

\section*{Acknowledgement}
We gratefully acknowledge the funding support received from the National Quantum Science and Technology Initiative (NQSTI), which made this research possible.

\end{document}